\documentclass[]{aa}

\usepackage{graphicx}
\usepackage[varg]{txfonts}
\pdfmapfile{+txfonts.map}
\usepackage{natbib}
\usepackage{array}
\usepackage{txfonts}
\usepackage{subfigure}
\usepackage{enumitem}
\usepackage{hyperref}
\usepackage{comment}
\usepackage{threeparttable,lscape}
\usepackage{multirow}
\usepackage{multicol}
\usepackage{float}
\usepackage{amsfonts}
\usepackage{amsmath}
\usepackage[title]{appendix}
\usepackage{multicol}
\usepackage{booktabs,multirow}
\usepackage{appendix}
\usepackage{comment}

\usepackage{xcolor}
\usepackage{pdflscape}
\usepackage{afterpage}

\begin{document}

   \title{Joint \textit{XMM-Newton} and \textit{NuSTAR} observations of the reflection spectrum of III\,Zw\,2}
 

   \author{Wara Chamani,
          \inst{1,2}\fnmsep\thanks{E-mail address: wara.chamani@aalto.fi}
          Karri Koljonen
          \inst{1,3} \fnmsep\thanks{E-mail address: karri.koljonen@utu.fi}
          \and
          Tuomas Savolainen\inst{1,2,4}
          }

   \institute{Aalto University Mets\"ahovi Radio Observatory, Mets\"ahovintie 114, FI-02540 Kylm\"al\"a, Finland 
        \and
             Aalto University Department of Electronics and Nanoengineering, PO Box 15500, FI-00076 Aalto, Finland 
         \and
             Finnish Centre for Astronomy with ESO(FINCA), University of Turku, V\"ais\"al\"antie 20, FI-21500 Piikki\"o, Finland
        \and
            Max-Planck-Institut f\"ur Radioastronomie,
            Auf dem H\"ugel 69, DE-53121 Bonn, Germany
        }

   \date{Received October 25, 2019; accepted February 10, 2020}

 
  \abstract
{Detecting and modelling the reprocessed hard X-ray emission component in the accretion flow, so-called reflection spectrum is a main tool to estimate black hole spins in a wide range of astrophysical black holes regardless of their mass or distance. In this work, we studied the X-ray spectra of the Seyfert~I galaxy III\,Zw\,2 using multi-epoch \textit{XMM-Newton}, \textit{NuSTAR} and \textit{Suzaku} observations. The X-ray spectra exhibit a soft-excess below 1\,keV and a
prominent excess at the location of the broad Fe\,K$\alpha$ line at 6.4 keV. To account for these spectral features, we have fitted the spectra with multiple models including an ionized partially covering absorber and an accretion disk reflection model. To fully resolve the reflection component, we analyzed jointly the \textit{XMM-Newton} and \textit{NuSTAR} observations taken in 2017 and archival \textit{XMM-Newton} data from 2000. Assuming the reflection scenario, the resulting model fits support a rapidly spinning black hole (\textit{a} $\geqslant$ 0.98) in this radio-intermediate source. The X-ray spectra in 2000 and 2017 are remarkably similar with the only difference in the reflection fraction, possibly due to a change in the geometry of the accretion flow. However, the \textit{Suzaku} observation is markedly different, and we suggest this could be an effect of a jet contribution in the X-ray band, which is supported by the elevated radio flux during this observation.
}

   \authorrunning{Chamani et al.}

   \maketitle
%

\section{Introduction}

\begin{table*}[t]
\centering
\caption[]{X-ray observations of III\,Zw\,2 used in this paper. The exposures shown in the table represent the total time on source.}
\label{t1}
\begin{tabular}{ccccccc}
\hline
\noalign{\smallskip}
Epoch & Mission/Instrument & OBSID & \begin{tabular}[c]{@{}c@{}} Date\end{tabular} & Exposure (ksec) & Total counts & Energy band (keV) \\
\noalign{\smallskip}\hline\noalign{\smallskip}
\multirow{2}{*}{2017} & \textit{NuSTAR}: FPMA/FPMB & 60301014002 & \multirow{2}{*}{2017-12-11} & 85.8/85.5 &21212/20626 & 3 - 79  \\
\noalign{\smallskip}
 & \textit{XMM-Newton}: EPIC-pn  & 0795620101 &  & 26.2 & 203720 & 0.23 - 10\\
 \noalign{\smallskip}
2011 & \textit{Suzaku}: XIS0/XIS3 & 706031010 & 2011-06-14 & 65.3/65.3 & 39067/41781 &0.7 - 10   \\
\noalign{\smallskip}
2000 & \textit{XMM-Newton}: EPIC-pn & 0127110201 & 2000-07-03 & 10.2 & 33264 & 0.23 - 10 \\
\noalign{\smallskip}\hline\noalign{\smallskip}
\end{tabular}
\end{table*}

Black hole (BH) spins are fundamental physical parameters that preserve information about the formation, evolution and the mass accretion rate history of supermassive black holes (SMBHs) residing in the centres of galaxies \citep{Volonteri2005, Volonteri2010}. The determination of the BH spin has been addressed by three different X-ray data analysis methods (for further details see \cite{Reynolds2008}). However, the majority of the spin measurements in active galactic nuclei (AGN) have been constrained by fitting the X-ray spectra with a model including a reprocessed (`reflected') component from the accretion disk. The reflection spectrum encodes information about the magnitude of the spin via the location of the innermost stable circular orbit (ISCO). Up-to-date, there are more than two dozen BH spin measurements with robust results, and a large number of them have high spin values \citep{Reynolds2019}. Due to the clear observational features the reflection method is one of the most preferred BH spin estimation methods, although there are some observational challenges that could affect the constraining of the spin, such as obtaining a sufficiently high signal-to-noise X-ray spectrum, assessing the effect of absorption and emission features on the continuum as well as weakness of the inner disk Fe\,K$\alpha$ line in sources seen at low inclination angles \citep{Laor2019}. 

The magnitude of the spin of SMBHs in AGN plays an important role in constraining relativistic jet production models. Magnetic fields can extract the rotational energy of a spinning BH via the Blandford-Znajek mechanism to launch jets \citep{Blandford1977}. The so-called “spin paradigm” attempts to explain the wide radio-loudness distribution \citep{Sikora2007} solely in terms of differences in the BH spin \citep{Blandford1999, WilsonColbert1995,Tchekhovskoy2010}. The existence of highly-spinning black holes in radio-quiet AGN does not fit well this paradigm \citep[e.g.,][]{Walton2013, Reynolds2014}. Hence, the production of powerful relativistic jets might not depend exclusively on high spins but also on the role of another parameter in the Blandford-Znajek mechanism, namely magnetic flux threading the black hole. For instance, \cite{Sikora2013} have proposed that the accretion history of a BH may determine whether high enough magnetic flux is accumulated to the BH to launch powerful jets. This latter model is sometimes called the “magnetic flux paradigm”. The magnetic flux, together with the BH spin determines the jet power -- hence the wide range of jet powers would not only depend on the range of spins but also on the accretion history of the source. The most powerful jets, which have power comparable to the accretion power of the system, can be explained by the magnetically arrested disk \citep[MAD;][]{Narayan2003, Tchekhovskoy2011, McKinney2012} scenario, in which the magnetic flux threading a BH is at the maximum value allowed by the accretion rate. 

The study of the distribution of BH spins in radio-quiet and radio-intermediate AGN can potentially help to distinguish between the two paradigms. Flat-spectrum radio-intermediate quasars (RIQ) are especially interesting sources in this respect, since it has been suggested that they are beamed counterparts of radio-quiet quasars \citep{Falcke1996a,Falcke1996b} and since their compact radio cores in principle allow the jet magnetic flux to be measured with very long baseline interferometry (VLBI) by using the core-shift effect \citep[e.g.,][]{Pushkarev2012, Zamaninasab2014}.

III\,Zw\,2 (also known as Mrk\,1501 and PG\,0007+106; \linebreak $z=0.089$) is a radio-intermediate source in the Palomar-Green (PG) survey of quasars \citep{Schmidt1983} but with a luminosity corresponding to a Seyfert galaxy type I \citep{Khachikian1974}. It hosts a supermassive black hole with a mass on the order of $10^8$$-$$10^9$\,M$_{\odot}$ \citep{Salvi2002, Hao2005, Vestergaard2006, Grier2012} and Eddington ratio of $\sim$0.1 \citep{Inoue2007}. A study for a period of 25 years has shown significant variability (an order of magnitude) from radio to X-ray wavelengths \citep{Salvi2002}. Recently, two gamma-ray flaring episodes have been detected by \textit{Fermi}-LAT that correlate well with optical and radio emission \citep{Liao2016}. High-resolution radio observations of III\,Zw\,2 have shown a compact radio core with a flat spectrum and a relativistic, although not very highly relativistic, jet that moves at an apparent superluminal speed of 1.3$-$2.7c \citep{Falcke1999, Brunthaler2000, Brunthaler2005, Lister2019}. In addition, III\,Zw\,2 has a weak extended radio emission \citep{Brunthaler2005} that continues to the kilo-parsec scale \citep{Cooper2007}. 

Previous X-ray spectral studies of III\,Zw\,2 have reported a significant, broad Fe\,K$\alpha$ line emission in the EPIC spectra of the \textit{XMM-Newton} observation in 2000 \citep{Salvi2002, Piconcelli2005, Jimenez2005}. \cite{Salvi2002} found that the best-fit model includes a reflection component ({\sc pexrav}) with an iron line at 6.45\,keV (EW$\sim$0.87\,keV). Further studies on the Fe\,K$\alpha$ line emission have been performed by \cite{Gonzalez2018} based on the archival data from \textit{XMM-Newton} in 2000 and \textit{Suzaku} in 2011. They modelled the spectra with a reflection component ({\sc xillver}; \cite{Garcia2010, Garcia2011, Garcia2013}) and provided interpretations to the nature of the reflector emitting the line. Based on the best fit results, they interpreted that the \textit{XMM-Newton} spectrum can be explained by reflection from a highly-ionized accretion disk while the \textit{Suzaku} spectrum is more consistent with a reflection from a cold torus. A highly spinning black hole has been assumed throughout their study. However, no spin estimate based on the observations has previously been reported for III\,Zw\,2.

In this work, we aim to estimate for the first time the BH spin using the X-ray spectra of III\,Zw\,2. In Section~2, we present our new X-ray observations from \textit{XMM-Newton} and \textit{NuSTAR} missions, archival data from \textit{XMM} and \textit{Suzaku}, and the corresponding data reduction procedures to obtain the calibrated data products. In Section~3, we describe the results of the spectral analysis that includes the model fitting of the data with phenomenological models to estimate the photon indices and black body temperature as well as testing a warm absorber model. Subsequently, we fit a model including a reflection component to two data sets: the \textit{XMM-Newton} observation of 2000 and our joint broadband \textit{XMM-Newton}+\textit{NuSTAR} observations of 2017. We have modeled these spectra with a reflection model {\sc relxill} \citep{Dauser2010, Garcia2014, Dauser2014, Dauser2016}, which allows us to resolve the reflection spectrum fully, and to give an estimate of the BH spin. In Section~4, we summarize our main results, which include the observation of a soft excess component on the \textit{XMM}-2000 and 2017 data sets as well as the Fe\,K$\alpha$ line profile. Jointly fitting the combined 2000 and 2017 spectra with {\sc relxill} suggests that III\,Zw\,2 harbours a fast-rotating black hole. In addition, we discuss some of the caveats of this work as well as future prospects.

\begin{figure}
  \centering
    \includegraphics[width=0.52\textwidth]{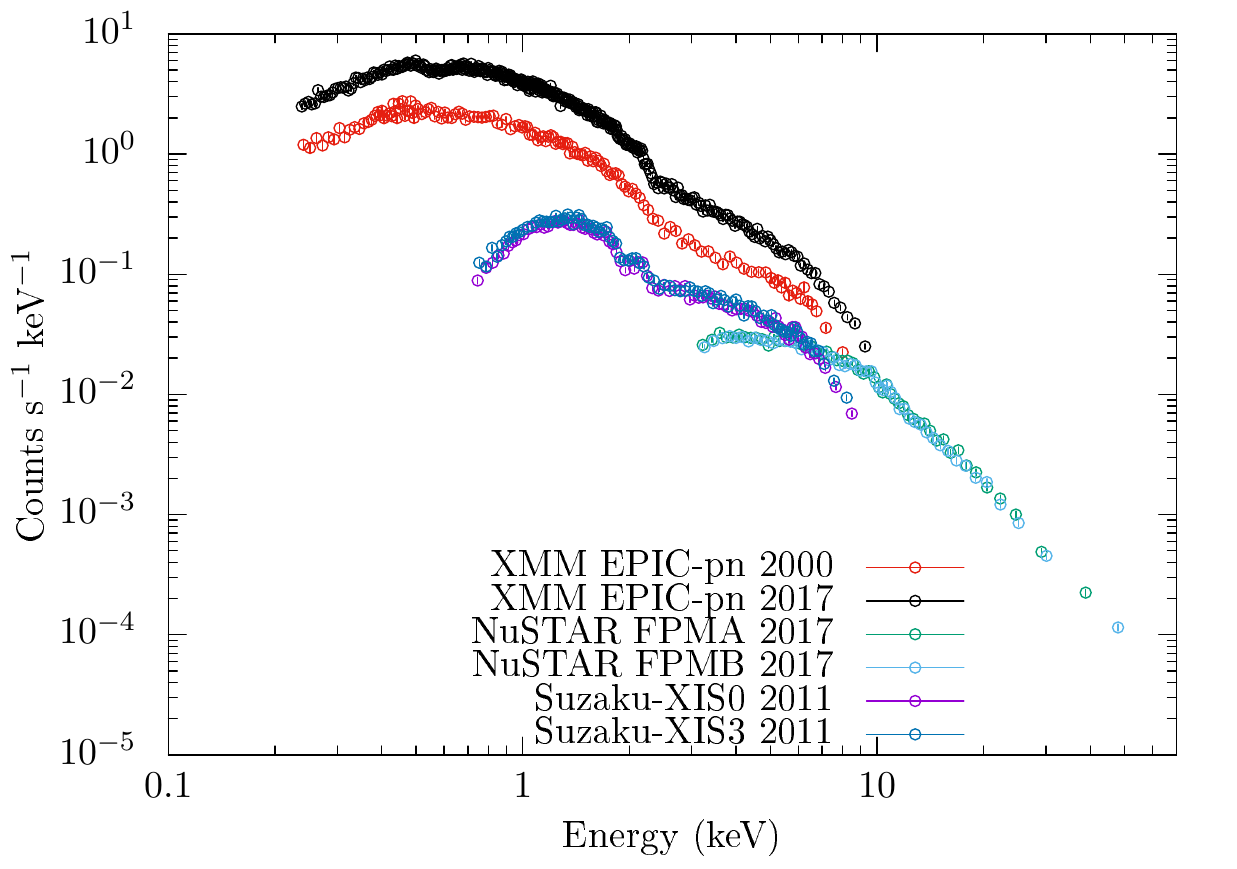}
    \caption{Multi-epoch folded spectra of III Zw 2 for the observations in 2000, 2011 and 2017.}
    \label{f1}
\end{figure}

\afterpage{
\clearpage
\renewcommand{\tabcolsep}{3pt}
\begin{landscape}
\centering
\begin{threeparttable}
\caption[]{The parameters for the best-fit models for all the datasets. Table columns are as follows: (1) the mission and the instrument used, (2) the year of the observation, (3) the fitted model, (4) the foreground hydrogen column density N$_\mathrm{H}$ in units of 10$^{22}$ cm$^{-2}$, (5)  the column density of the warm absorber $N_\mathrm{H,WA}$ in units of 10$^{22}$ cm$^{-2}$, (6) the ionization parameter of the warm absorber  $\xi'$ in units of erg cm s$^{-1}$, (7) the partial covering factor of the warm absorber \textit{f}, (8) the power law photon index $\Gamma$, (9) the black body temperature in units of eV, (9) the total, unabsorbed flux for each instrument in units of 10$^{-11}$ erg cm$^{-2}$ s$^{-1}$, (11) the fit quality shown as the ratio of the chi-square value $\chi^{2}$ over the degrees of freedom (d.o.f). The values in brackets represent the reduced chi-square value. }
\label{t2}
\begin{tabular}{lllllllllcl}
\hline
\noalign{\smallskip}
\multicolumn{1}{c}{\begin{tabular}[c]{@{}c@{}}Mission/Instrument\\ (1)\end{tabular}} & \multicolumn{1}{c}{\begin{tabular}[c]{@{}c@{}} Date\\ (2)\end{tabular}} & \multicolumn{1}{c}{\begin{tabular}[l]{@{}c@{}} Model\\ (3)\end{tabular}} &  \multicolumn{1}{c}{\begin{tabular}[c]{@{}c@{}} $N_\mathrm{H}\tnote{a}$  \\ (4)\end{tabular}} & \multicolumn{1}{c}{\begin{tabular}[c]{@{}c@{}} $N_\mathrm{H,WA}$ \\ (5)\end{tabular}} & \begin{tabular}[c]{@{}c@{}}$\log(\xi'$)\\ (6) \end{tabular} & \multicolumn{1}{c}{\begin{tabular}[c]{@{}c@{}} \textit{f}\\ (7)\end{tabular}}  & \multicolumn{1}{c}{\begin{tabular}[c]{@{}c@{}} $\Gamma$\\ (8)\end{tabular}}   & \multicolumn{1}{c}{\begin{tabular}[c]{@{}c@{}} $kT$ \\ (9)\end{tabular}} & \multicolumn{1}{c}{\begin{tabular}[c]{@{}c@{}} Total Flux \tnote{b}\\ (10)\end{tabular}} & \multicolumn{1}{c}{\begin{tabular}[c]{@{}c@{}}  $\chi^{2}$/d.o.f\\ (11)\end{tabular}} \\
\noalign{\smallskip}\hline\noalign{\smallskip}
\textit{XMM-Newton}/EPIC-pn & 2000 & PL & 0.060 $\pm$ 0.004 & - &- &- & 1.77 $\pm$ 0.02 & - & 5.70 & 259.1/139 (1.89) \\
&   & WA*PL & 0.105$^{+0.012}_{-0.011}$ &55$^{+48}_{-11}$  & 2.08$^{+0.09}_{-0.07}$ & 0.60$^{+0.05}_{-0.06}$ & 1.99$^{+0.06}_{-0.05}$ & - & 8.69 & 144.9/136 (1.07) \\
 \noalign{\smallskip}
&   & PL+BB & 0.083$_{-0.012}^{+0.013}$ &- &- &- & 1.67 $\pm$ 0.04 & \multicolumn{1}{l}{136$^{+15}_{-12}$} & 5.01 & 165.3/137 (1.21) \\
 \noalign{\smallskip}
&   & WA*(PL+BB) & 0.102$^{+0.017}_{-0.013}$  & 65$^{+102}_{-25}$  & 2.10$^{+0.67}_{-0.46}$ & 0.50$^{+0.22}_{-0.10}$  & 1.86$^{+0.12}_{-0.07}$ & 137$^{+43}_{-34}$  & 6.96 & 133.8/134 (0.99)\\
\noalign{\smallskip}\hline \noalign{\smallskip}
\textit{Suzaku}/XIS & 2011 & PL & 0.207 $\pm$ 0.011 &- & -&- & 1.58 $\pm$ 0.02 & - & 2.59 & 271.2/180 (1.51)\\
 \noalign{\smallskip} \hline \noalign{\smallskip}
\textit{XMM-Newton}/EPIC-pn & 2017 & PL &  0.069 $\pm$ 0.002 &- &- & -& 1.84 $\pm$ 0.01 & - & 14.2/9.3 & 1113.8/464 (2.40) \\
 \noalign{\smallskip}
\multicolumn{1}{c}{+} &  & WA*PL & 0.102$^{+0.004}_{-0}$ &48$^{+4}_{-3}$& 2.05$\pm$0.04  & 0.47$^{+0.02}_{-0.03}$ & 2.03$\pm$0.02 & -  & 19.6/9.5  & 683/461 (1.48) \\  
 \noalign{\smallskip}
\multicolumn{1}{c}{\textit{NuSTAR}} &  & PL+BB & 0.090$^{+0.005}_{-0.004}$ &- &- & -& 1.77 $\pm$ 0.01 & 144$^{+6}_{-5}$ & 12.6/9.3  & 531.6/462 (1.15)\\
  \noalign{\smallskip}
&  & WA*(PL+BB) & 0.101$\pm$0.006   & 14$^{+4}_{-3}$   & -2.18$^{+1.48}_{-0.82}$ & 0.19$^{+0.04}_{-0.05}$ & 1.86$^{+0.02}_{-0.03}$ & 130$^{+7}_{-5}$ & 14.7/9.3  &494.2/459 (1.07)\\
\noalign{\smallskip} \hline 
\end{tabular}
\begin{tablenotes}
\item [a] The Galactic value is N$_\mathrm{H}$ = 5.39 $\times$ $10^{20}$\,cm$^{-2}$ \citep{Kalberla2005}. 
\item [b] Absorption-corrected fluxes estimated for the following ranges: 0.23$-$10\,keV (2000), 0.7$-$10\,keV (2011). For the latter the total flux is the average of the XIS cameras. For the 2017 spectrum, separate calculations are made for the ranges: 0.23$-$10\,keV(\textit{XMM})/3$-$50\,keV(\textit{NuSTAR}). The total flux of the 3$-$50\,keV range is the average value of the two FPM detectors.

\end{tablenotes}
\end{threeparttable}
\end{landscape}
\clearpage
}

\section{Observations and data reduction}

\textit{NuSTAR} observations with the supporting \textit{XMM-Newton} observations are currently the only set of instruments that provides the necessary sensitivity and bandpass to study the accretion processes of AGN in detail: the hard X-ray band of \textit{NuSTAR} to study the reflection features of the iron line region and Compton reflection hump, and the soft X-ray band of \textit{XMM-Newton} to support the study of the reflection features in the iron line region and the soft excess.

We observed III\,Zw\,2 simultaneously with \textit{XMM-Newton} and \textit{NuSTAR} on 2017 December 11. In Table \ref{t1}, \textit{XMM-Newton} exposures are given for the EPIC-pn camera for the 0.2$-$10\,keV band with a mean countrate of 8.74 $\pm$ 0.02 counts s$^{-1}$ (1000 s bin time). \textit{NuSTAR} exposures are given for the two focal plane modules (FPM) in the 3$-$79\,keV band. The mean countrates for the 7000-s bin lightcurves for both detectors are  0.590 $\pm$ 0.004 and 0.607 $\pm$ 0.004  counts s$^{-1}$.

\subsection{\textit{XMM-Newton}}

In this work, we analyzed only the EPIC-pn source spectra due to the higher sensitivity compared to the MOS cameras. The 2017 observation of III Zw 2 with the EPIC-PN camera was executed in the Prime Small window mode with a thin filter for a total duration of 41\,ksec and an exposure time of 26\,ksec.
To obtain the spectra, we have processed the EPIC-PN data using the HEASOFT package version 6.24 and the \textit{XMM-Newton} Scientific Analysis System (SAS version 17.0.0) and followed the guidelines on the mission website\footnote{\url{https://xmm-tools.cosmos.esa.int/external/xmm_user_support/documentation/sas_usg/USG/}}. High-level science products have been obtained using the \textsc{xmmextractor} tool. For the light curve and spectral extraction, large background flaring was subtracted with the task TABGTIGEN by creating a good time interval file with a background countrate threshold of 0.4 count/s. No pile-up corrections are applied as the countrate during the observation is well below the small window pile-up threshold (25 counts/s). The energy band used for the spectral analysis ranges from 0.23 to 10 keV. 
 
\subsection{\textit{NuSTAR}}

The calibration and product extraction of the \textit{NuSTAR} data were performed by following the standard guidelines\footnote{\url{https://heasarc.gsfc.nasa.gov/docs/nustar/analysis/nustar_quickstart_guide.pdf}}.
We used the NUPIPELINE script to obtain calibrated event files for the FPMA and FPMB detectors. Subsequently, we used the NUPRODUCTS script to extract the instrumental responses, the source spectra from a 30" circular area centered on the source, and the background spectra from a 60" circular area from a source-free region. We also include the correction of South-Atlantic anomaly (SAA) effect on the background rates. The exclusion of SAA passages has been done by using the parameters: SAACALC=3, SAAMODE=optimized and TENTACLE=yes. For the spectral fitting we used both FPM spectra in the energy range of 3-50 keV with a constant in the model to account for any differences in the normalization between the two detectors. 

\subsection{Archival X-ray data}

In addition, we analyzed all the available X-ray observations of III\,Zw\,2 conducted in the past in comparable resolution in the soft X-ray domain, namely an \textit{XMM-Newton} observation from 2000 and a \textit{Suzaku} observation from 2011 (see Table~\ref{t1}). We reduced the \textit{XMM} data of the 2000 observation similarly to the 2017 data (see above). For the \textit{Suzaku} 2011 observation, we used the available data products from the XIS0 and XIS3 CCD cameras. \\

\subsection{Radio data}

The 37\,GHz radio observations of III\,Zw\,2 covering  1986$-$2019 were made with the 13.7\,m diameter Mets\"ahovi radio telescope in Finland (see Figure~\ref{f6} which shows the radio flux curve from 1997 to 2019). The detection limit of the telescope at 37\,GHz is on the order of 0.2\,Jy under optimal conditions. Data points with a signal-to-noise ratio < 4 are handled as non-detections. The flux density scale is set by observations of DR\,21. Sources NGC\,7027, 3C\,274 and 3C\,84 are used as secondary calibrators. A detailed description of the data reduction and analysis is given in \citet{Terasranta1998}. The error estimate in the flux density includes the contribution from the measurement rms and the uncertainty of the absolute calibration.

\begin{figure*}[t!]
    \centering
    \subfigure[ ]
    {
        \includegraphics[width=0.48\textwidth]{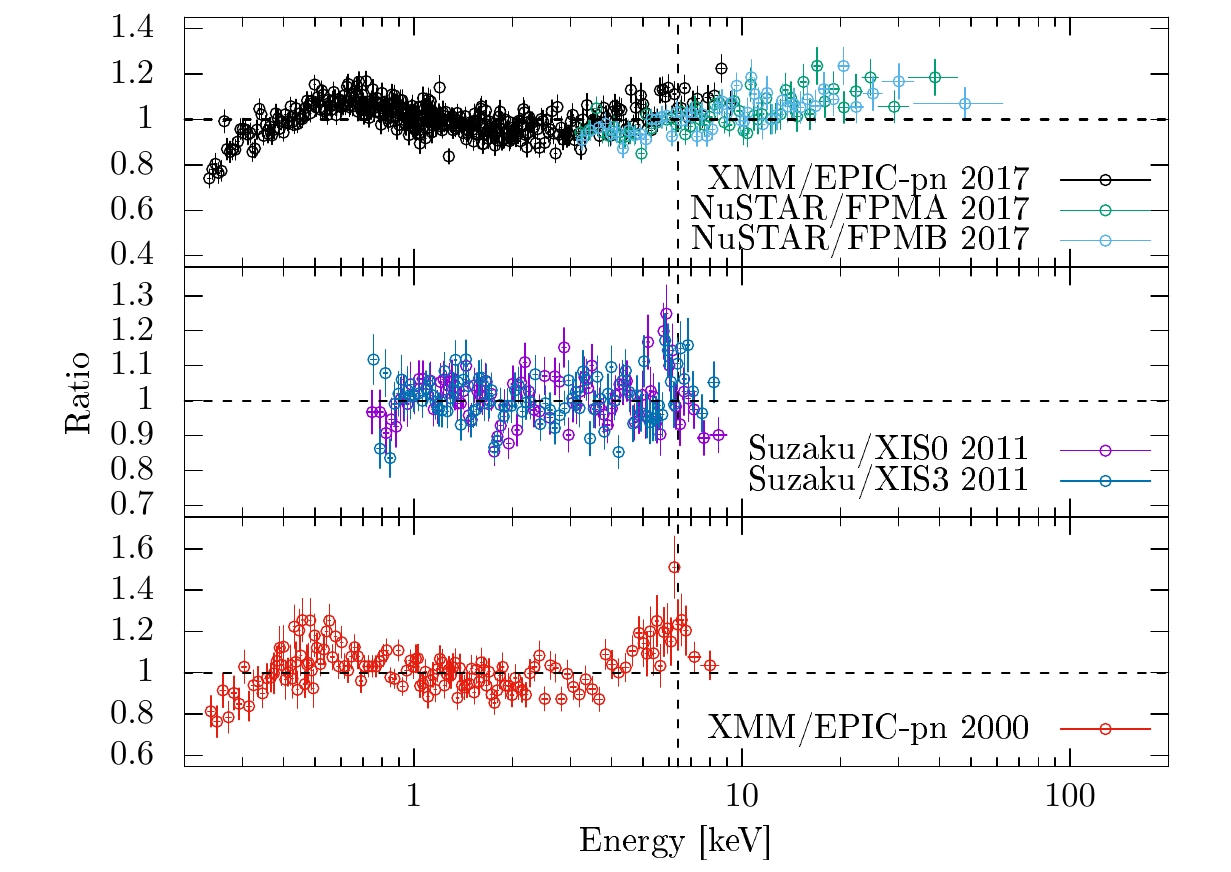}
    }
    \subfigure[]
    {
        \includegraphics[width=0.48\textwidth]{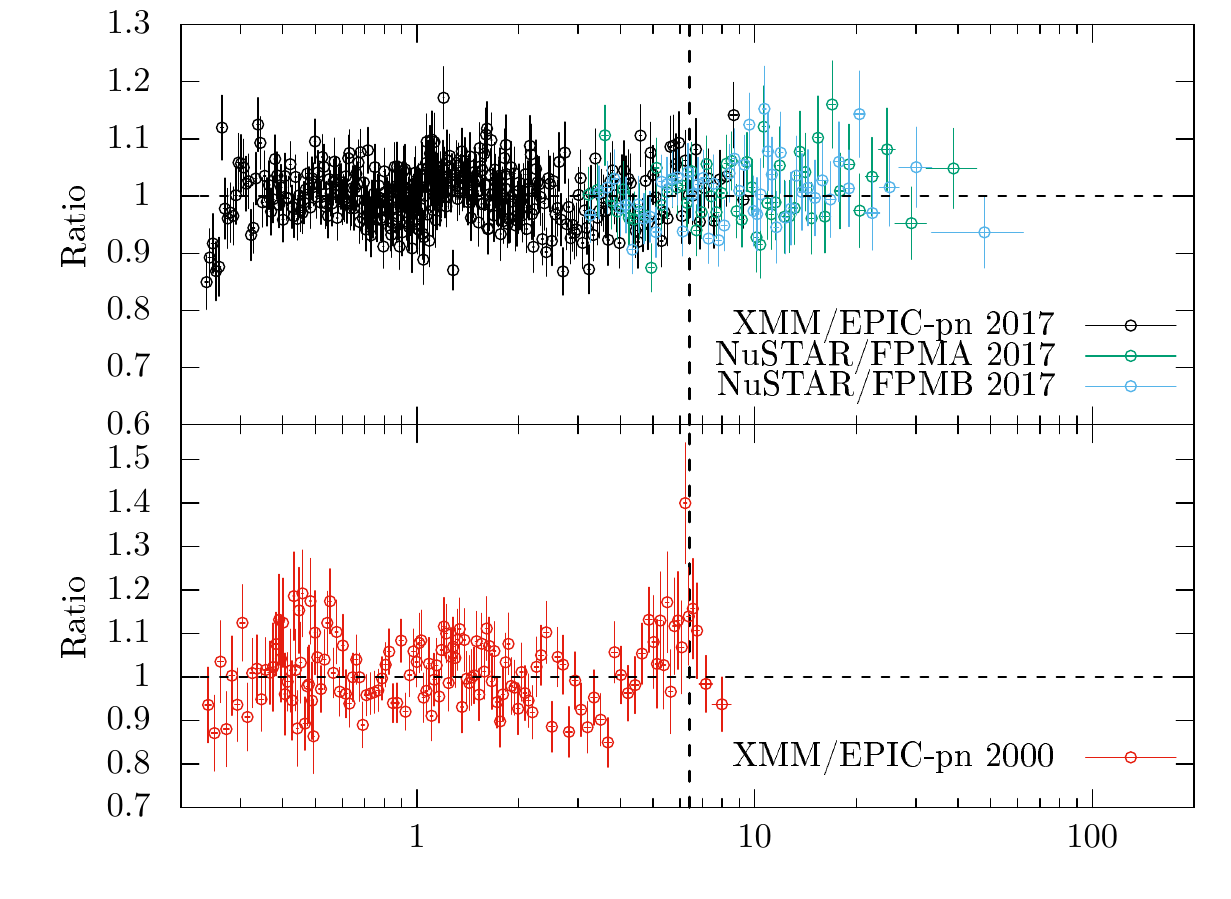}
    }
    \caption{The data-to-model ratio of (a) an absorbed power-law model and (b) an absorbed power-law+black-body radiation model for all epochs. The vertical dashed-line marks the position of the Fe K$\alpha$ line.}
    \label{f2}

\end{figure*} 


\begin{figure*}[t!]
    \centering
    \subfigure[ ]
    {
        \includegraphics[width=0.48\textwidth]{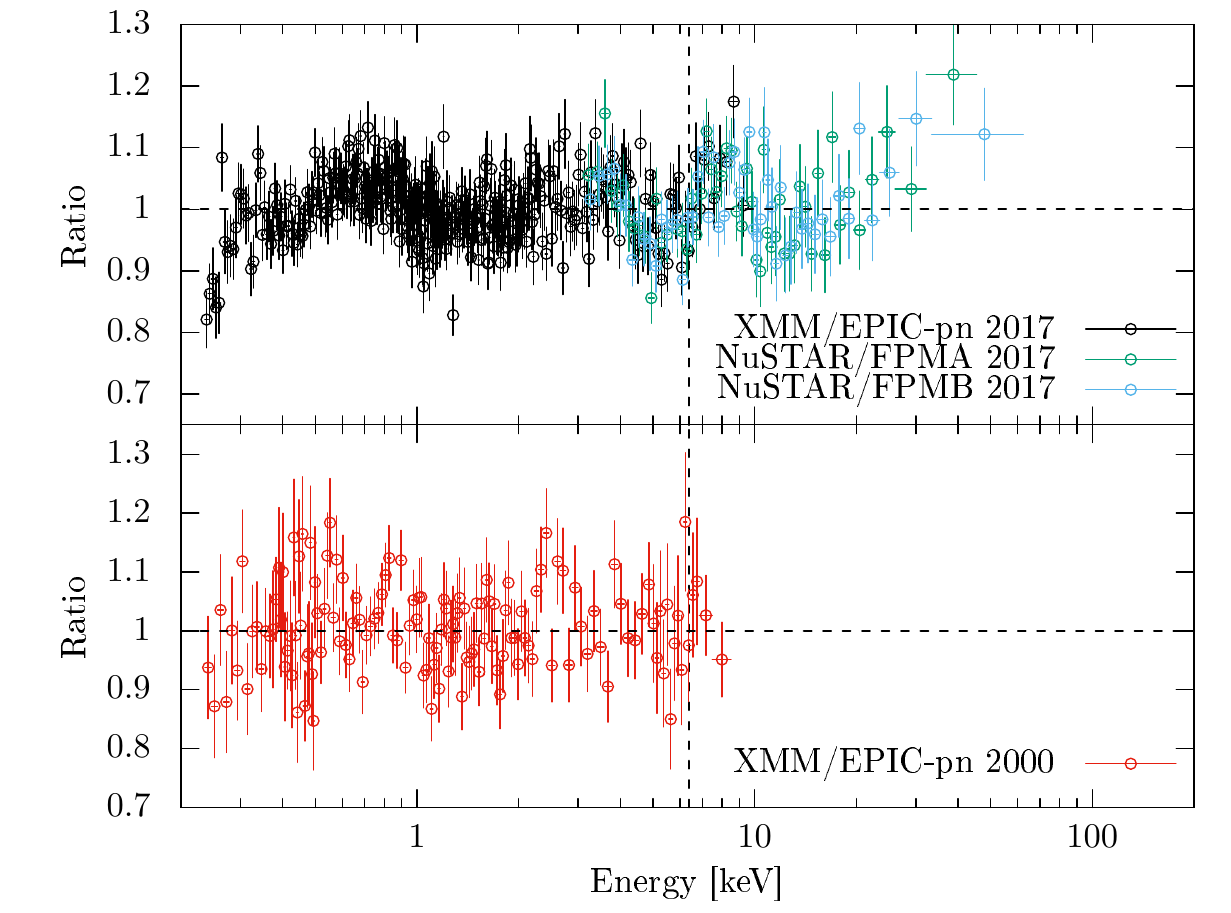}
    }
    \subfigure[]
    {
        \includegraphics[width=0.48\textwidth]{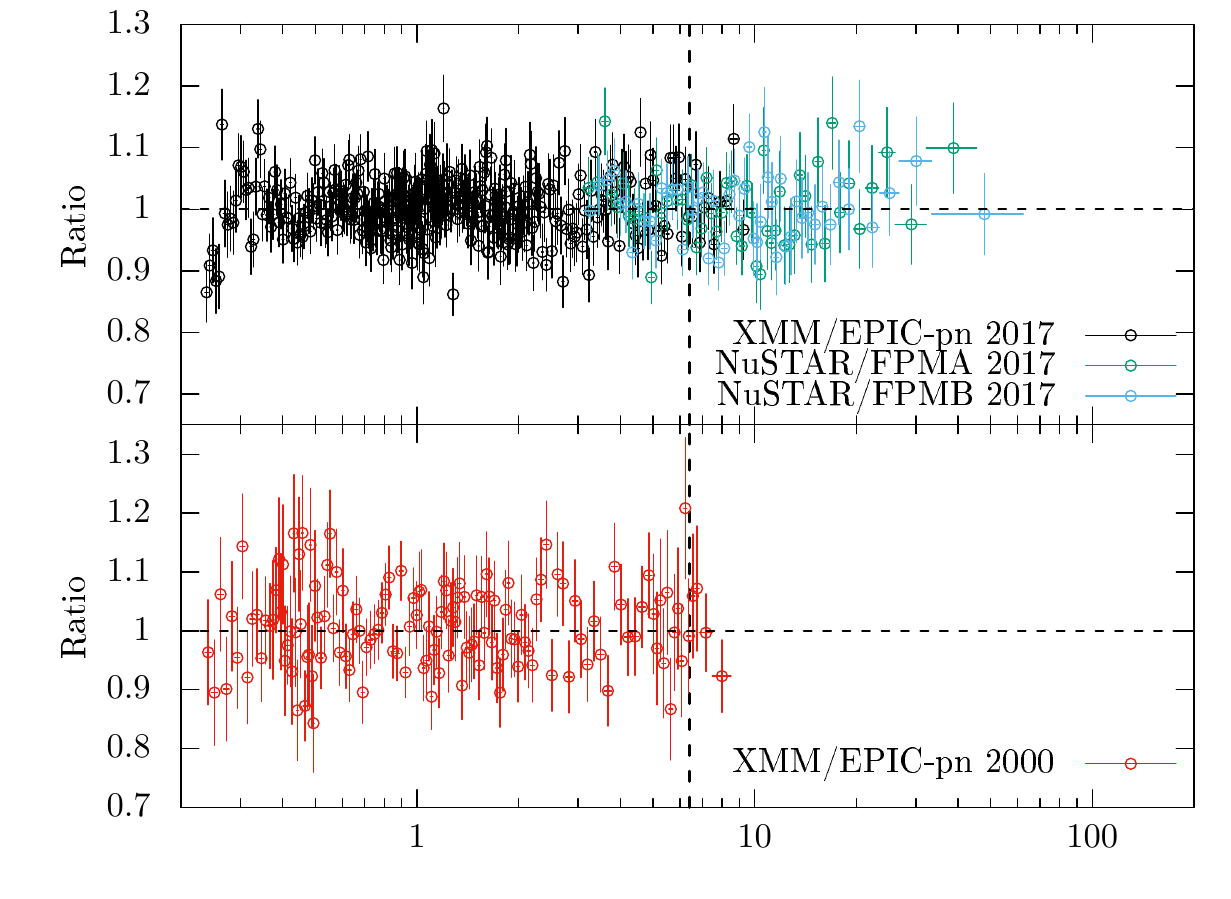}
    }
    \caption{The data-to-model ratio of (a) a power-law continuum model with partial covering of a partially ionised absorber and (b) the same as in (a) but with black body emission included.}
      \label{f3}
\end{figure*}

\begin{table*}
\centering
\begin{threeparttable}
\caption[]{{\sc relxill} and {\sc relxilllp} (i.e. lamp-post) model parameters. The fits shown in this table were done separately for the 2000 and 2017 datasets. The fixed parameters in the model are $\beta_1 = 5$, $\beta_2 = 3$, $R_\mathrm{br} = 4$\,$R_{g}$, $R_\mathrm{in}=1$($R_\mathrm{ISCO}$), $R_\mathrm{out} = 400$\,$R_{g}$ (see text for details).}
 \label{t3}
\begin{tabular}{llcc}
\hline\noalign{\smallskip}
Model     & Model parameter         & \textit{XMM}-2000 & Joint \textit{XMM}+\textit{NuSTAR} 2017 \\ 
\hline\noalign{\smallskip}\noalign{\smallskip}
RELXILL   & $N_\mathrm{H}$ &        0.097$^{+0.012}_{-0.016}$ & 0.102$^{+0.005}_{-0.003}$      \\ \noalign{\smallskip}
          & Normalization $\times$ 10$^{-5}$ & 3.44$^{+1.06}_{-0.76}$ & 8.15$^{+0.25}_{-0.46}$ \\ \noalign{\smallskip}
          & $\Gamma$ & 1.74$^{+0.05}_{-0.06}$ & 1.87 $\pm$ 0.01        \\ \noalign{\smallskip}
          & Total Flux\tnote{a} & 7.54 & 18.3/9.2 \\ \noalign{\smallskip}
          & \textit{a} & $\geqslant$ 0.78 &  $\geqslant$ 0.98     \\  \noalign{\smallskip}
          & $\theta$\tnote{b} & 38$^\circ$ $^{+3^{\circ}}_{-12^{\circ}}$ &  41$^\circ$ $^{+0^{\circ}}_{-1.5^{\circ}}$ \\  \noalign{\smallskip}
          & log($\xi$) & 2.75$^{+0.17}_{-0.27}$ & 2.70$^{+0.03}_{-0.06}$  \\ \noalign{\smallskip}
          & $A_\mathrm{Fe}$ (Solar) & 2.51 $^{+0.96}_{-1.44}$ & 2.11$^{+0.53}_{-0.69}$ \\  \noalign{\smallskip}
          & \textit{R} & 0.81 $\pm$ 0.41 & 0.49$^{+0.06}_{-0.05}$ \\  \noalign{\smallskip}
          & $\chi^{2}$/d.o.f & 136.7/134 (1.02) & 557.7/459 (1.21)  \\  \noalign{\smallskip}
RELXILL$\_$LP & $N_\mathrm{H}$ & 0.089 $\pm$ 0.011 & 0.098 $\pm$ 0.003 \\  \noalign{\smallskip}
          & Normalization $\times$ 10$^{-4}$ & 2.21$^{+0.65}_{-1.05}$ & 5.00$^{+0.10}_{-.0.07}$ \\ \noalign{\smallskip}
          & $\Gamma$   & 1.71$^{+0.05}_{-0.04}$ & 1.86 $\pm$ 0.01 \\  \noalign{\smallskip}
          & Total Flux\tnote{a}  & 7.49  & 19.3/9.8  \\  \noalign{\smallskip}
          & \textit{a} & $\geqslant$ 0.64 &  $\geqslant$ 0.97  \\ \noalign{\smallskip}
          & $\theta$\tnote{b}& 38$^\circ$ $^{+3^{\circ}}_{-9^{\circ}}$ & 41$^{\circ}$ $^{+0^{\circ}}_{-1^{\circ}}$ \\  \noalign{\smallskip}
          & log($\xi$) & 2.77$^{+0.18}_{-0.17}$ &  2.70$^{+0.02}_{-0.05}$ \\ \noalign{\smallskip}
          & $A_\mathrm{Fe}$ & 3.07$^{+0.78}_{-1.75}$ &  1.66$^{+0.51}_{-0.62}$  \\ \noalign{\smallskip}
          & \textit{R}& 2.71$^{+1.16}_{-1.42}$ & 1.60$^{+0.10}_{-0.21}$  \\  \noalign{\smallskip}
          & h\tnote{c} & $\leqslant 4.6$ & 3 \\  \noalign{\smallskip}
          & $\chi^{2}$/d.o.f & 137.3/133 (1.03) & 587.6/458 (1.28)    \\ \hline  \noalign{\smallskip}
\end{tabular}
\begin{tablenotes}
\item Note: N$_\mathrm{H}$ is given in units of 10$^{22}$ cm$^{-2}$, the power law normalization factor with the units of photons/keV/cm$^2$/s at 1 keV,  the total flux for each energy range in units of 10$^{-11}$ erg cm$^{-2}$ s$^{-1}$,  $\xi'$ in units of erg cm s$^{-1}$, $A_\mathrm{Fe}$ in solar units and h in R$_{g}$.
\item [a] Absorption-corrected fluxes estimated for the following ranges: 0.23$-$10\,keV (2000), 0.7$-$10\,keV (2011). For the latter the total flux is the average of the XIS cameras. For the 2017 spectrum, separate calculations are made for the ranges: 0.23$-$10\,keV(\textit{XMM})/3$-$50\,keV(\textit{NuSTAR}). The total flux of the 3$-$50\,keV range is the average value of the FPM detectors.
\item [b] The maximum disk inclination angle is set to 41$^\circ$.
\item [c] The minimum default value for the height is 3\,R$_{g}$.
\end{tablenotes}
\end{threeparttable}
\end{table*}

 
\section{Spectral analysis and results}
This work is devoted mainly to the spectral analysis of the X-ray data by fitting models to the observed source spectra for the epochs listed in Table~\ref{t1}. The analysis of the X-ray spectra was performed with ISIS\footnote{\url{https://space.mit.edu/asc/isis/docs.html}} (Interactive Spectral Interpretation System, package version 1.6.2-41; \citealt{Houck2000}). Before fitting we grouped the spectra such that each energy bin has S/N $\geq$10. The fit quality is estimated using $\chi^2$ statistics and the uncertainties associated with the model parameters are shown at the 90 per cent confidence level. 

Figure~\ref{f1} displays the X-ray spectra of III\,Zw\,2 from the \textit{XMM}-2000, \textit{Suzaku}-2011 and \textit{XMM+NuSTAR}-2017 observations. Overall, the spectral shape stays similar with a clear change in the flux level between the 2000 and 2017 observations. On the contrary, the \textit{Suzaku}-2011 spectrum shows a change in the spectral shape with a different power-law index and stronger absorption.

We have first fitted the spectra with a phenomenological model consisting of an absorbed power-law component (\textsc{phabs}$\times${\textsc{powerlaw}}; hereafter PL). We have not fixed the value of the neutral hydrogen column density ($N_\mathrm{H}$) as it was done in previous studies \citep{Salvi2002, Jimenez2005, Piconcelli2005, Gonzalez2018} in order to allow its possible variability over time \citep[the Galactic value is $N_\mathrm{H} \sim 5 \times 10^{20}$\,cm$^{-2}$;][]{Kalberla2005} and to take into account any variability in an additional, intrinsic absorption component.

Figure~\ref{f2}a shows the data/model ratio (residuals) of the PL model in the range 0.2$-$10\,keV for all epochs. A prominent soft X-ray excess (0.2$-$1\,keV) is observed in the \textit{XMM}-2017 spectrum similar to the 2000 data \citep{Gonzalez2018}. The existence of a soft excess in the \textit{Suzaku}-2011 spectrum is difficult to ascertain due to a flatter spectrum, higher absorption and a limited low energy response. The \textit{Suzaku}-2011 spectrum appears to be significantly more absorbed with the N${_\mathrm{H}}$ differing by a factor of at least two compared the other epochs (see Figure~\ref{f1}). Additionally, the spectrum presents a lower photon index ($\Gamma \sim 1.58$).

A prominent excess between 5$-$7\,keV is observed in the \textit{XMM}-2000 residuals indicating a contribution from the Fe\,K$\alpha$ line \citep{Salvi2002, Jimenez2005, Gonzalez2018}. Interestingly, such an excess appears to be also present in the \textit{Suzaku}-2011 residuals, however, it is less prominent and statistically less significant. In contrast to these findings, the \textit{XMM+NuSTAR}-2017 residuals show no significant Fe\,K$\alpha$ line. 

Based on the reduced $\chi^2$ values shown in Table~\ref{t2}, it is clear that pure power-law models are not suitable to describe the \textit{XMM}/\textit{NuSTAR} data. We point out here that since the \textit{Suzaku}-2011 data has a limited low energy response, it cannot be used to place any constraints on the soft excess. Thus, we conclude that the PL model can sufficiently explain the \textit{Suzaku}-2011 spectrum and do not fit any subsequent models to the data.

Secondly, we added a black body component to model the soft excess (hereafter PL+BB). The residuals of these fits for all epochs are shown in Figure~\ref{f2}b. We find that PL+BB model fits much better the \textit{XMM+NuSTAR}-2017 and the \textit{XMM}-2000 datasets. 
The spectral fitting results are displayed in Table~\ref{t2}.

The black body temperatures of the best fitting PL+BB models range between 136$-$192\,eV, which are consistent with the values reported by \citet{Piconcelli2005} and \citet{Gierlinski2004}. These values are much higher than the maximum temperature of the standard accretion disk ($\sim$10\,eV), which means that we cannot attribute the black body component to the thermal emission from the accretion disk. Such a large temperature of the soft excess has also been reported for several other AGN, including radio-quiet PG quasars as well as narrow line Seyfert 1 galaxies (NLS1s) \citep[][and references therein]{Ai2010}. However, no correlation between the soft excess temperature and the black hole masses (as low as 10$^{6}$M$_{\odot}$) has been found for the large majority of the sources with the exception of radio-loud NLS1s \citep{Yuan2010}.

In the above we have shown that a black body(-like) component is needed to fit well the spectrum of III\,Zw\,2. The  physical interpretation of the soft excess is a long-standing issue which has been addressed by different models. These include for instance the Comptonization of accretion disk photons to soft X-ray energies \citep[][and references therein]{Ai2010}, the reflection and absorption of photons in partially ionised material \citep{DoneNa2007, Done2007} and magnetic reconnection on the surface of the disk accelerating non-thermal electrons which then scatter photons via inverse Compton to soft-excess energies \citep{Zhong2013}. 

Phenomenological models including a black body component serve as a crude proxy for these physical scenarios and as such we cannot rule them out. However, testing individually a variety of such models is beyond the scope of this paper, since our goal is to constrain the spin value of the central black hole in III\,Zw\,2. In the rest of the paper we thus concentrate on modelling the spectra within a relativistic reflection scenario that can naturally explain both the soft excess and the excess around 6.4\,keV. For completeness, we first shortly discuss a warm absorber model, which is a commonly proposed alternative to reflection models for explaining the observed AGN X-ray spectra.

\subsection{Warm absorber model}

Even if the spectra of III\,Zw\,2 do not exhibit any strong absorption edges when compared to other sources \citep[see][]{Porquet2004}, we also investigate a scenario where so-called warm absorber (WA; model \textsc{zxipcf}) might be present. Thus, we have fitted the data with a model where partially ionised absorbing material partially covers the power-law continuum source (hereafter WA*PL) as well as with the same model including a black body emitter (hereafter WA*(PL+BB)). \\

The results of the fits are shown in Table~\ref{t2}. These models produce an improvement in the fit statistics compared to the PL and PL+BB models for the \textit{XMM}-2000 spectrum. While both models can effectively fit the excess around 6.4 keV, they do not fit well the soft excess (see residuals in Figures~\ref{f3}a,b; bottom panels). As for the 2017 spectrum, the WA*PL model produces only a slight improvement over the PL model and the residuals show a clear pattern in the soft X-rays similar to the PL model fit (Figure~\ref{f3}a; upper panel). The WA*(PL+BB) model produces a better fit, but essentially it is very similar to the PL+BB model (with some additional reprocessing in a cool gas), so the WA cannot be the physical interpretation to explain the soft excess in this case. The implied black-body temperature is again high and comparable to the values obtained previously without a warm absorber (see Table~\ref{t2}). We have also tried to fit the spectra with a power-law model modified by two warm absorbers, however, this model does not produce a significant improvement on the previous single warm absorber models fitted to the 2000 spectrum, i.e. we find similar fit quality. While a double warm absorber model can improve the single warm absorber power-law fit (with $\chi^2$/d.o.f = 589.4/458) to the 2017 spectrum, it does not fit well the data above 30\,keV.

Considering that the 2000 and 2017 spectra have on overall similar spectral shape, albeit with a different X-ray flux, and that there is a sign of a broad Fe\,K$\alpha$ line in the 2000 spectrum, in the next section we investigate a disk reflection scenario to estimate the black hole spin of III Zw 2. As we will see, the reflection model can explain both the soft excess emission and the blurred Fe\,K$\alpha$ line within a single physical scenario. 


\begin{table*}
\centering
\begin{threeparttable}
\caption[]{{\sc relxill} joint-fit parameters of the combined \textit{XMM}-2000 and \textit{XMM}+\textit{NuSTAR} 2017 spectra}.
\label{t4}
\begin{tabular}{cclll}
\hline
\noalign{\smallskip}
\multicolumn{5}{c}{Linked parameters}   \\ 
\hline
\noalign{\smallskip}
\multicolumn{1}{c}{$\theta$} & \multicolumn{1}{c}{\textit{a}} & \multicolumn{1}{c}{\textbf{N${_H}$}} & log($\xi$) & $A_\mathrm{Fe}$ \\ \noalign{\smallskip}\hline\noalign{\smallskip}
41$^{\circ}$ $^{+0^{\circ}}_{-1^{\circ}}$ &  $\geqslant$ 0.98  & \multicolumn{1}{c}{0.10$^{+0.04}_{-0.03}$} & 2.70$^{+0.02}_{-0.05}$   & 2.0 $\pm$ 0.6      \\ 
  \noalign{\smallskip}\hline\noalign{\smallskip}
\multicolumn{5}{c}{Unlinked parameters} \\ \noalign{\smallskip}\hline\noalign{\smallskip}
\multicolumn{2}{c}{Epoch} & Normalization  & \multicolumn{1}{c}{$\Gamma$}  & \multicolumn{1}{c}{\textit{R}}      \\ \noalign{\smallskip}\hline\noalign{\smallskip}
\multicolumn{2}{c}{2000} & \multicolumn{1}{c}{3.32$^{+0.25}_{-0.26}$} & \multicolumn{1}{c}{1.77$^{+0.02}_{-0.03}$} & 0.92 $\pm$ 0.14      \\
\noalign{\smallskip}
\multicolumn{2}{c}{2017} & \multicolumn{1}{c}{7.25$^{+0.28}_{-0.36}$} & \multicolumn{1}{c}{1.87$^{+0.01}_{-0.02}$} & 0.48 $\pm$ 0.05 \\
\noalign{\smallskip}
\hline
\noalign{\smallskip}
\multicolumn{3}{c}{$\chi^{2}$/d.o.f} & \multicolumn{2}{c}{696.6/598 (1.16)}  \\ \noalign{\smallskip}\hline\noalign{\smallskip}
\end{tabular}
\begin{tablenotes}
\item Note: N$_\mathrm{H}$ is given in units of 10$^{22}$ cm$^{-2}$, $\xi'$ in units of erg cm s$^{-1}$, $A_\mathrm{Fe}$ in solar units and the power law normalization factor with the units of 10$^{-5}$ photons/keV/cm$^2$/s at 1 keV.
\end{tablenotes}
\end{threeparttable}
\end{table*}

\subsection{Reflection model}

To explore physical conditions around the supermassive black hole in III\,Zw\,2 we employ the relativistic reflection model, {\sc relxill}\footnote{\url{http://www.sternwarte.uni-erlangen.de/~dauser/research/relxill/}} \citep{Garcia2014, Dauser2014}, to fit the data. {\sc relxill} models the irradiation (in coronal geometry) of the accretion disk by an emitter with a broken power law spectrum and a radial emissivity profile. It combines the reprocessed emission from the disk with the relativistic broadening effects on the emission features. For high-quality data the model allows the estimation of a number of parameters such as the disk ionization parameter ($\xi$), the iron abundance ($A_\mathrm{Fe}$), reflection fraction ($R$), inclination angle ($\theta$) and the black hole spin ($a$). For the disk irradiation profile, we have fixed the power-law indices to $\beta_1=5$, $\beta_2=3$ and the break radius to $R_\mathrm{br}=4$\,$R_\mathrm{g}$, corresponding to a low height of the primary X-ray source from the disk \citep{Dauser2013}. The energy cutoff of the primary spectrum is set to 300\,keV. Additionally, the gas density in the accretion disk is assumed to be $10^{15}$\,cm$^{-3}$ \citep{Garcia2016}.

To constrain the spin, an assumption about the disk inclination limits has been made based on the results of previous works. It is well known that III\,Zw\,2 has a radio jet exhibiting superluminal motion \citep{Brunthaler2000, Lister2019}. The maximum jet viewing angle of 41 degrees has been estimated from the apparent superluminal motion by \citet{Brunthaler2000}. In addition to this limit, various other estimates of the jet inclination can be derived from radio observations. \citet{Hovatta2009} report a jet inclination angle of 35$^\circ$ based on the flux variability time scale in the Mets\"ahovi 37\,GHz monitoring data and the apparent jet speed measured by the MOJAVE Survey \citep{Lister2019}. \citet{Liodakis2018} used a similar technique with the 15\,GHz flux variability data from the Owens Valley Radio Observatory monitoring program and reported a jet inclination angle of 22.1$^\circ$ with lower and upper limits of 0.9$^\circ$ and 23.5$^\circ$, respectively. Furthermore, the one-sidedness of the jet in III\,Zw\,2 can be used to place an upper limit of 53$^\circ$ to the jet inclination based on the required ratio of the Doppler boosting factors for the jet and the counter-jet.\footnote{In 2004.4, there was an ejection of a new jet emission feature that had a speed of $\beta_\mathrm{app}=1.36$ \citep{Lister2019}. This jet feature is well-separated from the core in the MOJAVE 15\,GHz VLBA image taken on August 9, 2007. We have measured the surface brightness ratio between the approaching jet and the counter-jet at a distance of 1.2\,mas from the VLBI core. Since there is no detectable emission on the counter-jet side, we use three times the image rms noise of 0.23\,mJy/beam as an upper limit. The resulting lower limit to the brightness ratio $R_\mathrm{B}$ is 64 and the corresponding upper limit to the jet inclination is $\theta < \mathrm{arccot} (\frac{R_\mathrm{B}^{1/(k-\alpha)}-1}{2\beta_\mathrm{app}}) \approx 53^\circ$, where $k=3$ for a transient jet feature and spectral index $\alpha$ is assumed to have the canonical value of $-0.7$.} Taken together, the upper limit of 41$^\circ$ for the jet inclination appears well-motivated. We assume that the jet is launched perpendicular to the accretion disk and thus set the maximum limit for the disk inclination also to 41 degrees. 

We first fitted the 2000 and 2017 data sets individually, then jointly. Results of the individual fits are shown in Table \ref{t3}. The confidence limits are calculated simultaneously for all free model parameters. We find that the {\sc relxill} models produce a similar fit quality as the warm absorber models for the 2000 data set and the black body component models for the 2017 data set.


Assuming that the reflection component causes the soft excess, we obtain a very large spin for the 2017 dataset. Due to the large spin and a low reflection fraction, the modelled iron line is heavily broadened and rather inconspicuous in the 2017 dataset, while in the 2000 data set the line is more visible due to increased reflection fraction (by a factor of two). We have also explored the scenario where a change of the location of the illuminating source to the accretion disk explains the time variable spectral features. For this we have fitted the data with the lamp-post scenario ({\sc relxilllp}). In this model, we allow the height ($h$) of the primary radiating source to vary freely. However, we find no significant differences in the heights between the two epochs. 
The lower limits to the spin values produced by either {\sc relxill} or {\sc relxilllp} models are significantly different for the first epoch. 

We have also attempted fitting the \textit{XMM}-2000 and \textit{XMM+NuSTAR}-2017 data sets with the reflection model affected by a warm absorber (hereafter WA*{\sc relxill}). The 2000 spectrum can be well fitted ($\chi^2$/d.o.f=134.2/131) with this model similar to the non-WA reflection model, so we do not find any significant improvement on the fits. The lower limit of the spin is, however, affected when including the warm absorber in the fit of \textit{XMM}-2000 data ($a \geq 0.46$). While the WA*{\sc relxill} can model well the soft excess of the 2017 spectrum, it fails to fit the hard X-ray component (>30\,keV). This is in accordance with the results by \citet{Risaliti2013}, who found for a much higher SNR spectrum of NGC\,1365 that partial covering absorption models can effectively fit the data up to 10\,keV; but deviate from the data at higher energies.

In addition, we have performed a joint fit to the combined 2000 and 2017 data sets with the {\sc relxill} model. A joint-fit consists of linked and unlinked parameters. Linked parameters are values which we assume do not change in human timescales such as the BH spin, the disk inclination, the ionization parameter and the iron abundance. Since we have not found any significant differences in the neutral hydrogen column density for the two \textit{XMM}-epochs (see Table~\ref{t2}), we let this to be also a linked parameter. The unlinked parameters such as the power-law photon index, normalization and the reflection fraction were allowed to vary independently. 

\begin{figure*}[t!]
   \centering
   \includegraphics[width=0.50\textwidth]{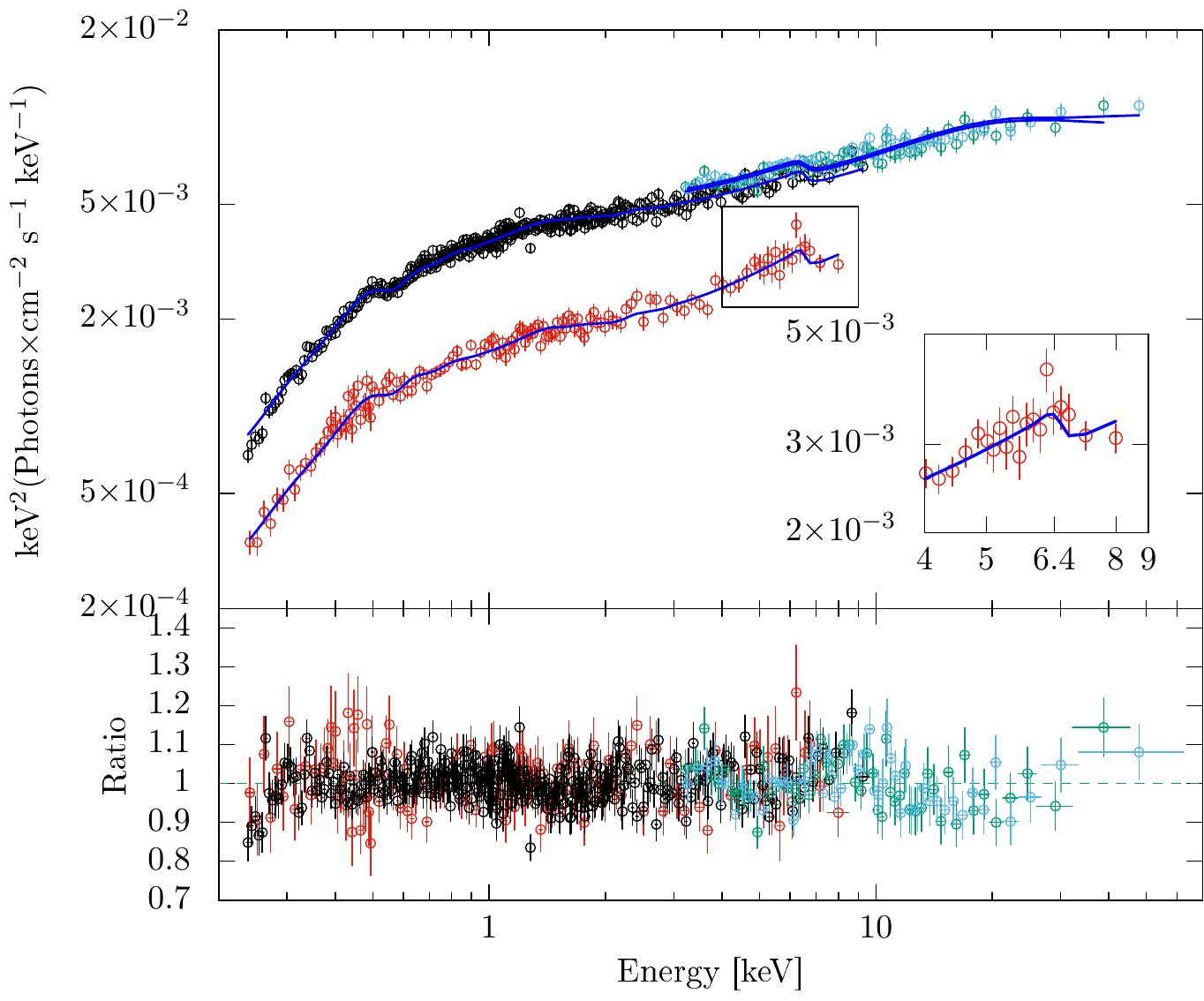}
   \caption{The joint-fit to the combined 2000 and 2017 data sets with the {\sc relxill} model (blue line). The data-to-model ratio is shown in the lower panel. The model fits the iron line as a heavily broadened and asymmetric feature indicating emission from very close to the black hole event horizon.}
   \label{f5}%
\end{figure*}

The best fit produces a nearly maximal spin of $a \geq 0.98$ with the disk inclination pegged to the maximum allowed value of 41 degrees (see Table~\ref{t4}). Additionally, the values of the other parameters appear to be similar to the values obtained in Table \ref{t3}. The joint-fit model to the multi-epoch spectra is shown in Figure~\ref{f5}. We note here that we have also studied cases, where the inclination angle is fixed to lower values (see Table \ref{ta1}.1 ) such as 22 degrees  \citep{Liodakis2018} and 35 degrees \citep{Hovatta2009} with a clear impact being to the value of the iron abundance, the ionization parameter and the reflection fraction. As a sanity check we have also fitted the data allowing the irradiation profile parameters to vary freely. The model converges to steep power-laws ($\beta_1=8.8$ and $\beta_2=3.1$ with $R_\mathrm{br}=2.8$\,$R_\mathrm{g}$) as is expected if the reflection comes from the inner accretion disk. The increase in the number of free parameters leads to a fit that is statistically slightly better ($\chi^2/\mathrm{d.o.f.}=669.5/595$) but does not properly reproduce the hard X-ray part of the spectrum. Therefore, we favor the fit with the irradiation profile kept fixed. 

We note that the 
{\sc relxill} model is able to fit well both spectra with only varying the reflection factor, power-law index and normalization. On the other hand, as shown in Figure~\ref{t1}, the X-ray flux has clearly increased in the 0.23$-$10\,keV range over 17 years. One possible explanation for the increased X-ray luminosity is the change in the mass accretion rate by a factor of $\sim$2. 

Since the best fit to the combined data set in Table~\ref{t4} gives an inclination at the upper edge of the allowed range, we have extended our analysis by fitting the {\sc relxill} model to the data also with a wide range of assumed disk inclinations. The results of these fits to the combined 2000+2017 data set are shown in Appendix~A (see Table \ref{ta1}.1). By letting the inclination vary freely, it settles on $\theta = 55^\circ \pm 2^\circ$. For this model the black hole spin is higher than 0.99. The overall fit quality as a function of inclination shows a clear valley around $\sim$50$^\circ$. Both low ($\lesssim$20$^\circ$) and high ($\gtrsim$60$^\circ$) inclination cases produce a poor fit to the iron line.

Finally, we extended the analysis to examine the quality of the fit for a wide range of spin values. The results are displayed in Table \ref{b1}. The minimum $\chi^{2}$ value is found at the very high (positive) black hole spin ($a=0.992$). 
Overall, we find a lower fit quality when the spin is set to values $\leq$ 0.9. In addition, low positive and all negative spin values show a strong tendency towards very large disk inclinations (hitting the upper limit of 80$^\circ$), sub-solar iron abundances, and low reflection fractions for both epochs. 
Given the limits to the jet inclination angle derived from the radio observations, and the fact that III\,Zw\,2 is a type 1 AGN with prominent broad emission lines \citep{Marziani2003}, we consider it rather unlikely that the disk inclination would be so high. We discuss this further in the next section.

\section{Summary and discussion}

We have studied the X-ray spectrum of III\,Zw\,2 for three different X-ray observations made in 2000, 2011 and 2017. To explore the presence of a soft X-ray component and an Fe\,K$\alpha$ line in the data, we have applied first phenomenological power-law models. These results have shown visible soft-excess component in the 2000 \citep{Gonzalez2018} and 2017 data sets. In addition, a prominent iron line has been observed in the \textit{XMM}-2000 data \citep{Salvi2002,Piconcelli2005,Gonzalez2018}. We studied whether the soft excess and the iron line can be fitted with a reflection model.

The black hole spin of III\,Zw\,2  has been constrained by applying the relativistic reflection model {\sc relxill}. We have estimated the BH spin using two different methods: by fitting the data separately and by combining the two observation epochs (2000 and 2017) under the joint-fit approach. In both methods, we have set an upper limit of 41$^\circ$ for the disk inclination. Fitting the 2000 spectrum alone produces a lower limit for the spin of $a \geqslant 0.78$, while the 2017 spectrum gives $a \geqslant 0.98$. Since the BH spin is not expected to change in a time scale of years, a joint-fit is considered as a preferable method to determine the spin of III\,Zw\,2. The joint fit favours a large spin value with a lower limit, $a\geqslant$ 0.98. We note that the quoted lower limits correspond to the statistical uncertainty of the fit and do not take into account any systematic effects. While systematics are difficult to assess, we can take the results of \citet{BonsonGallo2016} as a guideline. Those authors have studied the ability of the standard fitting procedure to recover the correct model parameters from a large number of simulated Seyfert 1 spectra. Their results indicate that spin can be measured with confidence to an accuracy of about $\pm0.1$ for high spin cases ($a > 0.8$) -- especially if the measured spectrum extends up to 70\,keV. Finally, we note that large spin values ($0.85 < a < 0.92$) have been also reported by \cite{Gnedin2012}. Their results are based on spectropolarimetric observations combined with the jet kinetic power and assuming a large disk inclination angle of $60^\circ$. \\

\begin{figure}[t!]
   \centering
   \includegraphics[width=0.48\textwidth]{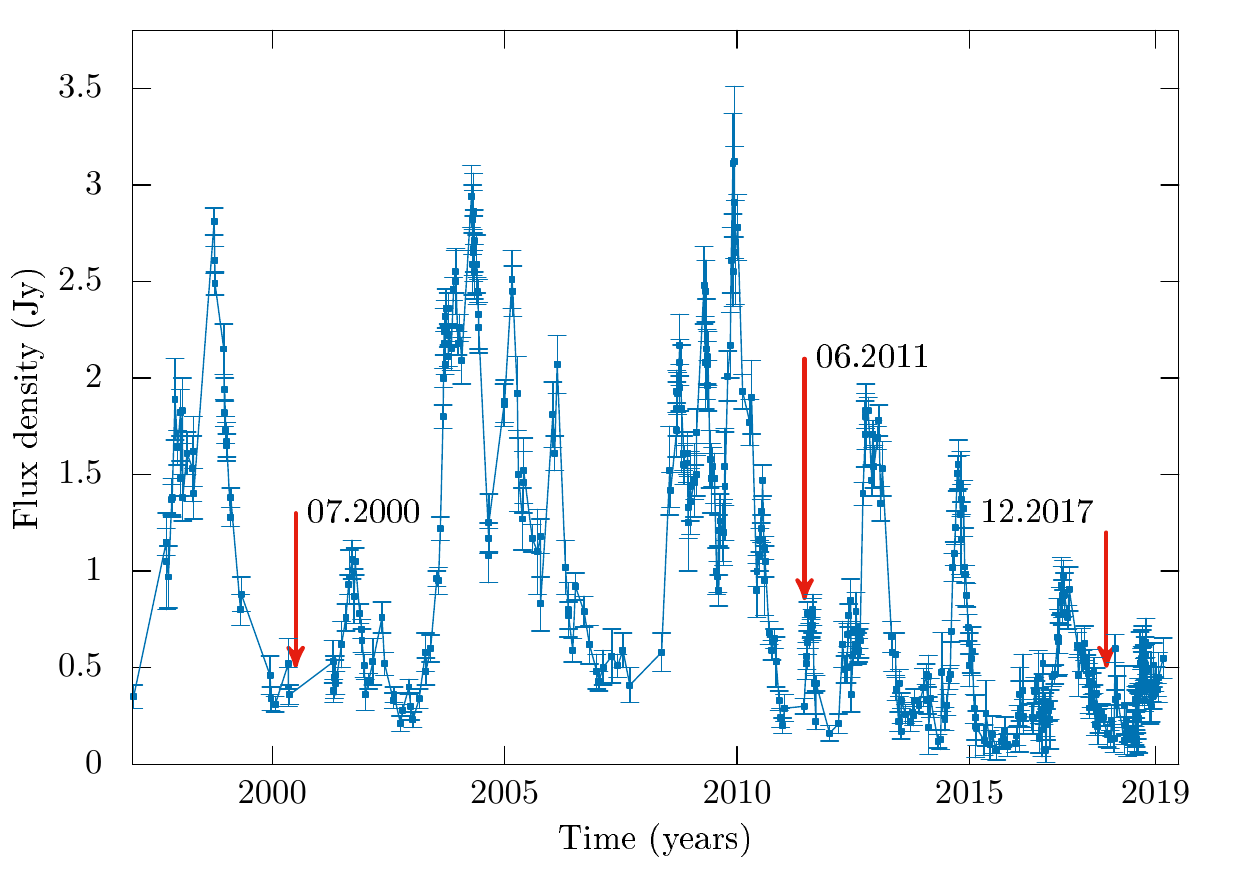}
      \caption{37\,GHz radio observations of III\,Zw\,2 obtained at the Mets\"ahovi Radio Observatory. The red arrows indicate the epochs when X-ray observations have been performed.}
      \label{f6}
\end{figure}

The estimation of the black hole spin by fitting a relativistic reflection model has naturally a number of caveats which we discuss below:
   \begin{itemize}
      \item It is possible that the X-ray spectrum is affected by intervening partially ionized absorbing material. If this is the case, the magnitude and the shape of the reflected component would be different from what is assumed here depending on the amount of absorption and could affect to our results presented above. While we cannot exclude such a warm absorber model, it requires an additional black body component of an unspecified origin in order to fit the data. Therefore, we favour here the reflection scenario as it is able to fit the data within a single physical scenario without adding phenomenological components.\\
   
      \item Since the black hole spin in our model fits is sensitive to the assumed disk inclination, it is important to have external constraints for it. We have assumed throughout this work that the accretion disk is perpendicular to the jet and the jet viewing angle is constant. The alignment of the jet with the angular momentum vector of the disk is the simplest assumption and it has some observational support from the measurements of the X-ray binary system XTE\,J1550$-$564 \citep{Steiner2012}. Recent general relativistic magneto-hydrodynamic simulations of tilted accretion disks around black holes have shown that the disk and the jet are generally well-aligned \citep{Liska2018}. However, Bardeen-Petterson effect can align the jet and the disk with the black hole spin axis at small radii (less than a few $R_\mathrm{g}$) rapidly compared to the viscous time scale in which the whole disk-jet system aligns with the black hole spin axis \citep{Liska2019}. This can temporarily leave the outer jet misaligned with respect to the inner accretion disk. It is not clear how large this misalignment can be, but for example, the simulations made by \citet{Liska2019b} indicate $\lesssim$15$^\circ$ difference even when the disk is originally tilted by 60$^\circ$ with respect to the black hole spin axis. Hence, while we cannot exclude the possibility that the inner disk inclination differs from the jet inclination, misalignment of tens of degrees seems unlikely.
      
      
      \item  The jet direction may not stay constant over time. While a precessing jet has been suggested for III\,Zw\,2 based on the quasi-periodic variations in its radio flux curves \citep{Li2010}, the measured jet kinematics rule out a large precession angle. Namely, there are three emission features which were ejected into the VLBI jet of III\,Zw\,2 during the past 20 years and which have robust kinematics measurements. All of them are moving in the same direction with the position angles of their motion differing less than $\sim$8$^\circ$ \citep{Brunthaler2000, Brunthaler2005, Lister2019}
      
      \item We find no prominent Fe\,K$\alpha$ line emission in the 2017 spectrum. This behaviour seems to be well explained by a heavy broadening of the iron line due to a high spin and a variable reflection fraction, which might be a signature of a change of the disk and/or corona geometry during a flare event or during a post-flare event leading to a decreased illumination of the disk. However, 37\,GHz radio observations of III\,Zw\,2 show that the source was in a very low radio state in 2000 and 2017. The epochs are highlighted in Figure~\ref{f6}. 
      
      \item We do not discard the possibility of X-ray emission from the jet, which is known to dominate the radio emission from III\,Zw\,2 \footnote{\citet{Liao2016} have even proposed that the spectral energy distribution of III\,Zw\,2 can be well-fitted by a single-zone, leptonic blazar model during the rapid $\gamma$-ray flare in 2009.}. This emission might have swamped the iron line and contaminated the continuum spectrum in 2017. However, we note that the power law photon indices in 2000 and 2017 model fits are on average close to 1.8, similar to what have been found in radio-quiet quasars \citep{Piconcelli2005} associated with the Comptonized hot corona of the disk \citep{Landt2008}. On the contrary, the spectrum in 2011 is highly absorbed with a photon index close to 1.6 that is similar to that of radio-loud quasars \citep{Piconcelli2005}. Thus, we suggest that the 2011 spectrum has an additional X-ray contribution from the jet most likely due to Synchrotron-Self Compton emission. In addition to this, it is interesting to note that \textit{Suzaku} observation in 2011 coincides with the rising part of a small radio flare at 37\,GHz while the 2000 and 2017 X-ray observations coincide with very low radio flux states that are indicative of low jet contribution (see Figure~\ref{f6}). Thus, we suggest here that in general a good strategy would be to perform X-ray observations when these types of sources are in their lowest radio state. This may indicate a suitable condition for the detection of the reflection component.
    
    \end{itemize} 
           
If the observed excesses below 1\,keV and at 5$-$7\,keV in the spectrum of III\,Zw\,2 are due to inner disk reflection, our modelling indicates a nearly maximally spinning black hole for this source. This result is the same whether using our initial assumption on the maximum disk inclination (41$^\circ$) or allowing the disk inclination to vary freely (see Table \ref{ta1}.1). While the limit of $a \geqslant 0.98$ corresponds to statistical uncertainties only and does not include systematic effects such as the finite thickness of the accretion disk \citep{Taylor2018}, our results do favour a fast-spinning BH in a source that belongs to a group of radio-intermediate quasars, which have been suggested to be relativistically beamed counterparts of radio-quiet quasars \citep{Falcke1996b}. If the jet in III\,Zw\,2 is indeed intrinsically weak, this suggests that spin is not the (only) parameter driving the vast differences in jet production efficiencies of accreting black holes.

Since the radio emission from the extended jet in III\,Zw\,2 is very weak compared to its beamed core emission \citep{Falcke1996a, Cooper2007} and since its optical emission is dominated by the accretion disk \citep{Chen2012}, we can estimate the intrinsic radio-loudness of III\,Zw\,2 by simply scaling the observed radio-loudness by $\delta^{-2.5}$, where $\delta$ is the relativistic Doppler factor and we have assumed a continuous jet with a spectral index of $-0.5$ \citep{Hovatta2014}. The range of reported radio-loudness in III\,Zw\,2 is 150$-$200 \citep{Falcke1996b,Sikora2007} and the Doppler factor estimates range from $\sim$2 to $\gtrsim$6 \citep{Hovatta2009,Liao2016,Liodakis2018}. This gives a rough estimate that the intrinsic radio-loudness in III\,Zw\,2 is in the range of 2$-$35, which indeed would place it in or close to the radio-quiet group.

The next step will be to measure the jet's magnetic flux with VLBI observations. The prediction of the "magnetic flux paradigm" is that III\,Zw\,2 -- with its fast spinning black hole -- should have magnetic flux well below the MAD-limit in order to explain the relatively low jet efficiency \citep{Sikora2013}.

\begin{acknowledgements}

We thank Stefanie Komossa for her valuable suggestions and comments during the writing of this paper. We also thank the anonymous referee for useful comments that helped to improve the manuscript. This work was supported by the Academy of Finland under the project "Physics of Black Hole -Powered Jets" (project numbers 274477, 284495, and 312496). This publication makes use of data obtained at the Mets\"ahovi Radio Observatory, operated by the Aalto University in Finland. Data and/or software have been provided by the High Energy Astrophysics Science Archive Research Center (HEASARC), which is a service of the Astrophysics Science Division at NASA/GSFC and the High Energy Astrophysics Division of the Smithsonian Astrophysical Observatory. This research has also made use of data obtained from the \textit{Suzaku} satellite, a collaborative mission between the space agencies of Japan (JAXA) and the USA (NASA). This research has made use of data from the MOJAVE database that is maintained by the MOJAVE team \citep{Lister2018}. K.I.I.K. was supported by the Academy of Finland project 320085.

\end{acknowledgements}


\bibliographystyle{aa}
\bibliography{references}

\onecolumn

\appendix
\appendixpage
\addappheadtotoc
\section{}

\begin{figure}[h!]
   \centering
   \includegraphics[width=0.55\textwidth]{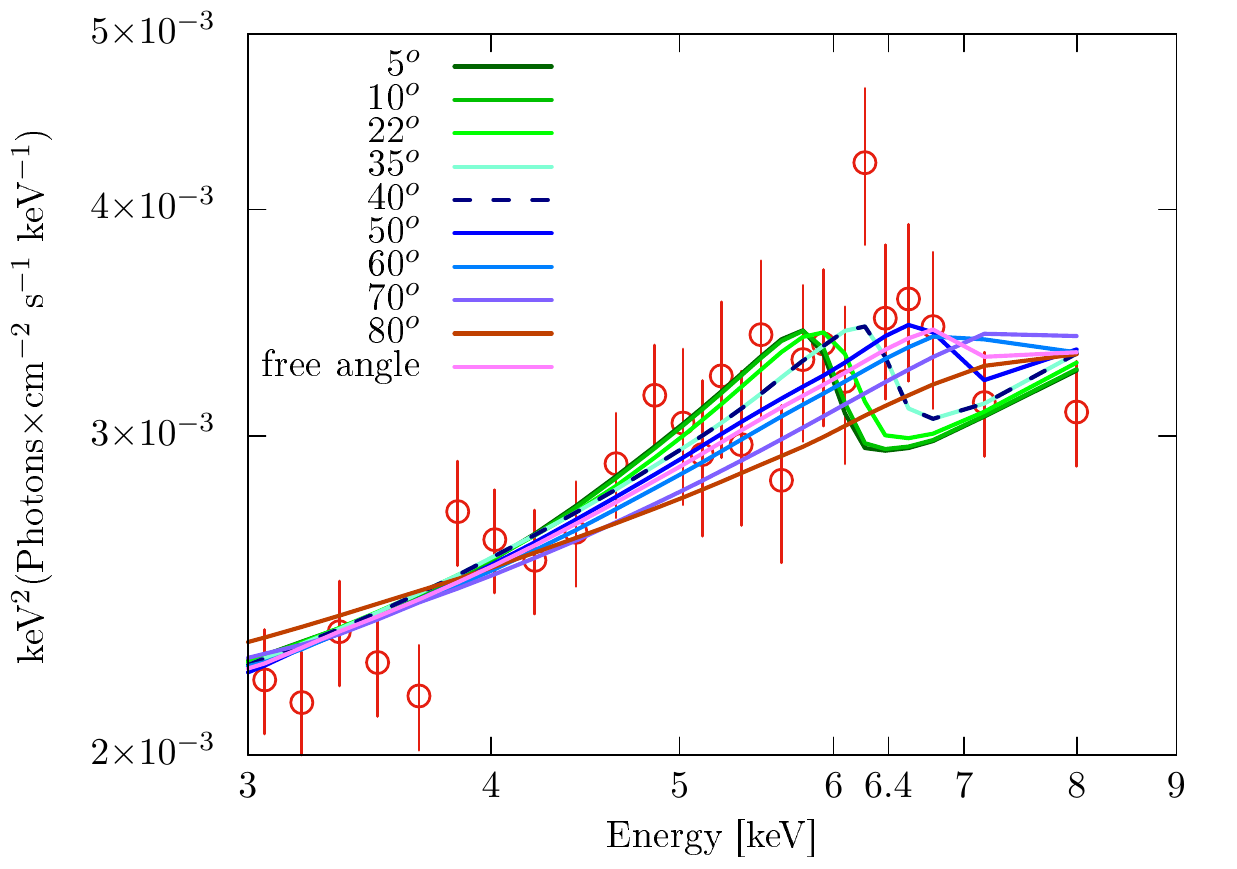}
      \caption{The {\sc relxill} fits to the high-energy edge of the iron line 
      of the combined 2000 and 2017 data sets for the different inclination angles shown in Table \ref{fa1}. The figure shows 3$-$9\,keV part of the spectrum. At lower angles, the model fits to the iron line are skewed towards the left side whereas at higher angles the fits display a linear tendency with a change of the slope. The models around the line resemble a broadened and asymmetric feature when the disk inclination lies for instance between 40 and 55 degrees.}
      \label{fa1}
\end{figure}

Below we show the best-fit parameters for {\sc relxill} models that have different fixed disk inclination values, as well as for the freely varying, non-constrained inclination case. The joint-fit with {\sc relxill} is given to the combined 2000 and 2017 spectra. We use the same fixed values for the irradiation profile and disk radius parameters as described in section 3.2. The joint-fits to the iron line region are displayed in Figure \ref{fa1}.

\begin{table}[!h]
\centering
\begin{threeparttable}
\label{ta1}
\caption[]{{\sc relxill} joint-fit results for different disk inclinations. The last row shows the results for a freely varying disk inclination.}
\begin{tabular}{lllllllll}
\hline
\noalign{\smallskip}
\multicolumn{1}{c}{$\theta$} & \multicolumn{1}{c}{\textit{a}} & 
\multicolumn{1}{c}{$\Gamma_{2000}$} &
\multicolumn{1}{c}{$\Gamma_{2017}$} &\multicolumn{1}{c}{\begin{tabular}[c]{@{}c@{}}log($\xi$)\\ (erg cm s$^{-1}$)\end{tabular}}& A$_\mathrm{Fe}$(solar) & \multicolumn{1}{c}{ \textit{R}$_{2000}$} & \multicolumn{1}{c}{\textit{R}$_{2017}$} & \multicolumn{1}{c}{$\chi^{2}$/d.o.f} \\
\noalign{\smallskip}\hline\noalign{\smallskip}
5$^\circ$ & $\geqslant$ 0.976 & 1.78$^{+0.02}_{-0.03}$ & 1.87$^{+0.02}_{-0.01}$ & 2.73$^{+0.04}_{-0.01}$ & 2.71$^{+0.50}_{-0.52}$ & 1.04$^{+0.18}_{-0.16}$ & 0.55$^{+0.07}_{-0.05}$ & 777/599 (1.30)\tnote{*} \\\noalign{\smallskip}
10$^\circ$ & $\geqslant$ 0.978 & 1.78$^{+0.02}_{-0.03}$ & 1.87$^{+0.02}_{-0.01}$  & 2.73$^{+0.04}_{-0.03}$  & 2.67$^{+0.50}_{-0.52}$ & 1.04$^{+0.18}_{-0.12}$ & 0.56$\pm$0.06 & 772/599 (1.29)\tnote{*}  \\\noalign{\smallskip}
22$^\circ$ &  $\geqslant$ 0.983  & 1.78$^{+0.02}_{-0.03}$ & 1.87$^{+0.02}_{-0.01}$ & 2.73$^{+0.03}_{-0.05}$ & 2.47$^{+0.51}_{-0.53}$ & 1.07$^{+0.18}_{-0.16}$ & 0.56$^{+0.06}_{-0.05}$ & 746.8/599 (1.25)\tnote{*}  \\\noalign{\smallskip}
35$^\circ$ & $\geqslant$ 0.986 & 1.77$\pm$0.03 & 1.87$^{+0.02}_{-0.01}$ & 2.71$^{+0.02}_{-0.05}$ & 2.18$^{+0.52}_{-0.56}$ & 0.99$^{+0.17}_{-0.14}$ & 0.52$^{+0.06}_{-0.05}$ & 710.4/599 (1.19) \\ \noalign{\smallskip}
40$^\circ$ & $\geqslant$ 0.988 & 1.77$^{+0.02}_{-0.03}$ & 1.87$\pm$0.01 & 2.70$^{+0.04}_{-0.05}$ & 2.05$^{+0.54}_{-0.56}$ & 0.93$^{+0.15}_{-0.14}$ & 0.49$\pm$0.05 & 698.3/599 (1.17) \\\noalign{\smallskip}
50$^\circ$ & $\geqslant$ 0.992 & 1.76$\pm$0.02 & 1.86$\pm$0.01  & 2.70$^{+0.02}_{-0.06}$ & 1.56$^{+0.54}_{-0.57}$ & 0.83$^{+0.15}_{-0.11}$ & 0.43$^{+0.05}_{-0.04}$ & 691.3/599 (1.15) \\\noalign{\smallskip}
60$^\circ$ & $\geqslant$ 0.993 & 1.73$\pm$0.02 & 1.84$\pm$ 0.01 & 2.70$^{+0.02}_{-0.07}$ & 1.19$^{+0.65}_{-0.23}$ & 0.56$^{+0.08}_{-0.09}$ & 0.31$\pm$0.03 & 693/599 (1.17)\tnote{*}  \\\noalign{\smallskip}
70$^\circ$ & $\geqslant$ 0.993 & 1.85$\pm$0.02 & 1.92$^{+0.02}_{-0.01}$ & 1.30$^{+0.08}_{-0.17}$ & 0.72$^{+0.23}_{-0.22}$ & 0.48$\pm$0.07 & 0.31$^{+0.03}_{-0.04}$ & 708/599 (1.18)\tnote{*} \\\noalign{\smallskip}
80$^\circ$ & $\leq$ 0.298 & 1.82$^{+0.01}_{-0.02}$  & 1.89$\pm$0.01  & 1.11$^{+0.20}_{-0.09}$  & 0.64$^{+0.26}_{-0.14}$  & 0.29$^{+0.01}_{-0.05}$  & 0.18$^{+0.03}_{-0.02}$  & 729.6/599 (1.22)\tnote{*} \\\noalign{\smallskip}
55$^\circ \pm 2^{\circ}$ & $\geqslant$ 0.992 & 1.74$\pm$0.02  & 1.85$\pm$0.01  & 2.70$^{+0.02}_{-0.05}$  & 1.57$^{+0.51}_{-0.46}$  & 0.64$^{+0.15}_{-0.09}$  & 0.35$^{+0.05}_{-0.03}$  & 685.2/598 (1.15) \\\noalign{\smallskip}
\noalign{\smallskip}\hline\noalign{\smallskip}
\end{tabular}
\begin{tablenotes}
\item [*] This model does not fit well the iron line. The plots exhibiting this effect are displayed in Figure \ref{fa1}.
\end{tablenotes}
\end{threeparttable}
\end{table}


\clearpage

\section{}
Below we show the {\sc relxill} fits to the combined 2000 and 2017 data sets. The best-fit parameters are displayed for different fixed spin values. The disk inclination and the other parameters vary freely.

\begin{table}[!h]
\centering
\caption[]{{\sc relxill} joint-fit results for different spin values.}
\label{b1}
\begin{tabular}{lllllll}
\hline
\noalign{\smallskip}
\textit{a} & $\theta$ &  log($\xi$) & A$_{Fe}$  & R$_{2000}$ & R$_{2017}$ & $\chi^{2}$/d.o.f 
\\ \noalign{\smallskip}\hline\noalign{\smallskip}
0.992 & 55$^\circ$ $^{+2^\circ}_{-3^\circ}$ & 2.70$^{+0.02}_{-0.07}$ & 1.35$^{+0.69}_{-0.37}$ & 0.64$^{+0.13}_{-0.11}$  & 0.35$\pm$0.04 & 688.3/599 (1.15)        \\ \noalign{\smallskip}
0.900 & 46$^\circ$ $^{+4^\circ}_{-6^\circ}$ & 2.70$^{+0.01}_{-0.05}$ & 1.29$^{+0.80}_{-0.31}$ & 0.59$^{+0.12}_{-0.08}$ & 0.30$\pm$0.04 & 726.6/599 (1.21)    \\ \noalign{\smallskip}
0.800 & 62$^\circ$$\pm$4$^\circ$  & 1.30$^{+0.08}_{-0.15}$ & 0.60$^{+0.22}_{-0.10}$ & 0.43$^{+0.08}_{-0.07}$ & 0.26$^{+0.03}_{-0.04}$ & 735.6/599 (1.23)   \\ \noalign{\smallskip}
0.700 & 62$^\circ$ $^{+7^\circ}_{-2^\circ}$ & 1.30$^{+0.05}_{-0.23}$ & 0.59$^{+0.23}_{-0.09}$ & 0.42$^{+0.06}_{-0.10}$ & 0.25$^{+0.03}_{-0.05}$ & 736.2/599 (1.23)       \\\noalign{\smallskip}
0.500 & 66$^\circ$ $^{+14^\circ}_{-5^\circ}$ & 1.29$^{+0.05}_{-0.55}$ & 0.60$^{+0.24}_{-0.10}$ & 0.39$^{+0.07}_{-0.11}$ & 0.23$^{+0.03}_{-0.05}$ & 733.9/599 (1.22)\\ \noalign{\smallskip}
0.300 & 78$^\circ$ $^{+2^\circ}_{-15^\circ}$ & 1.07$^{+0.25}_{-0.27}$ & 0.63$^{+0.26}_{-0.13}$ & 0.32$^{+0.10}_{-0.07}$ & 0.20$\pm$0.03 & 730.9/599 (1.22)  \\ \noalign{\smallskip}
0.100 & 78$^\circ$ $^{+2^\circ}_{-12^\circ}$ & 1.09$^{+0.23}_{-0.08}$ & 0.64$^{+0.25}_{-0.14}$ & 0.32$^{+0.09}_{-0.06}$ &   0.20$\pm$0.03 &728.6/599 (1.22)          \\ \noalign{\smallskip}
0  & 78$^\circ$ $^{+2^\circ}_{-11^\circ}$ & 1.11$^{+0.21}_{-0.09}$  & 0.64$^{+0.25}_{-0.14}$  & 0.32$^{+0.08}_{-0.06}$  & 0.20$\pm$0.03   &  727.8/599 (1.21)         \\ \noalign{\smallskip}
-0.100  & 78$^\circ$ $^{+2^\circ}_{-11^\circ}$ & 1.12$^{+0.19}_{-0.09}$ & 0.64$^{+0.25}_{-0.14}$ & 0.32$^{+0.08}_{-0.06}$ & 0.20$\pm$0.03  & 727.3/599 (1.21)         \\ \noalign{\smallskip}
-0.300  & 78$^\circ$ $^{+1^\circ}_{-10^\circ}$ & 1.15$^{+0.17}_{-0.11}$ & 0.65$^{+0.25}_{-0.15}$ & 0.32$^{+0.07}_{-0.06}$ & 0.20$^{+0.04}_{-0.03}$  & 726.7/599 (1.21)         \\ \noalign{\smallskip}
-0.500  & 78$^\circ$ $^{+1^\circ}_{-10^\circ}$ & 1.18$^{+0.15}_{-0.12}$  & 0.65$^{+0.25}_{-0.15}$  & 0.32$^{+0.08}_{-0.05}$  & 0.20$^{+0.04}_{-0.02}$   & 727.8/599 (1.21)         \\ \noalign{\smallskip}
-0.700 & 77$^\circ$ $^{+2^\circ}_{-8^\circ}$ & 1.22$^{+0.12}_{-0.14}$  & 0.66$^{+0.24}_{-0.16}$  & 0.32$^{+0.08}_{-0.05}$  & 0.20$^{+0.04}_{-0.02}$   & 727/599 (1.21)         \\ \noalign{\smallskip}
-0.800 & 77$^\circ$ $^{+2^\circ}_{-8^\circ}$ & 1.23$^{+0.12}_{-0.15}$  & 0.66$^{+0.24}_{-0.16}$  & 0.32$^{+0.08}_{-0.05}$  & 0.20$\pm$0.02   & 727.2/599 (1.21)         \\ \noalign{\smallskip}
-0.900 & 77$^\circ$ $^{+2^\circ}_{-8^\circ}$ & 1.25$^{+0.10}_{-0.16}$  & 0.66$^{+0.24}_{-0.16}$  & 0.32$^{+0.08}_{-0.05}$  & 0.20$\pm$0.02  & 727.6/599 (1.21)         \\ \noalign{\smallskip}
-0.992 & 77$^\circ$ $^{+2^\circ}_{-8^\circ}$ & 1.26$^{+0.09}_{-0.16}$  & 0.66$^{+0.24}_{-0.16}$  & 0.32$\pm$0.04  & 0.20$\pm$0.02  & 728/599 (1.22)         \\ \noalign{\smallskip}
 \hline
\noalign{\smallskip}
\end{tabular}
\end{table}

\end{document}